# Correlated Insulating States and Transport Signature of Superconductivity in Twisted Trilayer Graphene Moiré of Moiré Superlattices

Kan-Ting Tsai[1†], Xi Zhang[1†], Ziyan Zhu[2], Yujie Luo[1], Stephen Carr[2], Mitchell Luskin[3], Efthimios Kaxiras[2,4], Ke Wang[1]*

[1] School of Physics and Astronomy, University of Minnesota, Minneapolis, MN 55455, USA; [2] Department of Physics, Harvard University, Cambridge, MA 02138, USA; [3]School of Mathematics, University of Minnesota, Minneapolis, MN 55455, USA; [4] John A. Paulson School of Engineering and Applied Sciences, Harvard University, Cambridge, MA 02138, USA; † These authors contributed equally to this work; *To whom the correspondence should be addressed: kewang@umn.edu

Layers of two-dimensional materials stacked with a small twist-angle give rise to beating periodic patterns on a scale much larger than the original lattice, referred to as a "moiré superlattice." When the stacking involves more than two layers with independent twist angles between adjacent layers, it generates "moiré of moiré" superlattices[1], with multiple length scales that control the system's behavior. Here we demonstrate these effects of a high-order moiré superlattice in twisted trilayer graphene with two consecutive small twist angles. We report correlated insulating states near the half filling of the moiré of moiré superlattice at an extremely low carrier density (~ $10^{10}$ cm$^{-2}$), near which we also report a zero-resistance transport behavior typically expected in a 2D superconductor. Moreover, the temperature dependence of the measured resistances at full-occupancy ($v$ = -4 and $v$ = 4) states are semi-metallic, distinct from the insulating behavior of twisted bilayer systems[2–6], providing the first demonstration of emergent correlated transport behaviors from continuous, non-

**isolated higher-order moiré flat bands. Our findings shed new insights into the microscopic mechanisms of moiré correlated states and provide the impetus for future studies on this material platform, such as the demonstration of phase coherence and Meissner-like effect.**

When two layers of van der Waals (vdW) materials are placed on top of each other with a twist angle, isolated flat bands may emerge near the zero energy[7,8], giving rise to unconventional correlated phases including Mott insulator[3,9–11] states and superconductivity[2]. This discovery sparked extensive theoretical[12–23] and experimental investigations[24–27] into its microscopic mechanism, and recent efforts have been geared toward extensions to other material combination[28], including higher-order moiré systems consisting of multiple twisted pieces of vdW materials[1,29–32].

In this work, we fabricate a novel twisted trilayer graphene (tTLG) system with two independently controlled twist angles, in which three pieces of monolayer graphene are transferred on top of each other with designed twist angles in the range of 2-3° (fig. 1a) between adjacent layers. The full trilayer stack is encapsulated in hexagonal boron nitride (hBN), and bubble-free regions of a few μm in size are identified. The sample is then etched down to 3 μm × 3.5 μm size with a 1D edge contact defined at each corner (fig. 1a inset), allowing van der Pauw type 4-probe transport measurements. In this platform, a "moiré of moiré" (MoM) superlattice forms due to the interference of two sets of misaligned twisted bilayer superlattices between the 1-2 and 2-3 layer pairs (fig. 1b shows an visual example of the case of same interlayer twist angle 9.43°). Enhancement of the electronic density of states (DOS) occurs for a wide range of twist angles, at which the electron correlation may become important, making tTLG a unique platform to explore correlated states with more versatile experimental control. In this work, we report controlled experimental studies of two MoM lattices (defined in the following as MoM-1 and MoM-2), with unit cell sizes extracted to be $A_1$ = 6438 nm$^2$ and $A_2$ = 4778 nm$^2$, respectively (see later sections and SI for details). In contrast to the moiré superlattice in tBLG, a given unit cell size of MoM lattice in tTLG can correspond to a range of different possible twist angle combinations, that is, the MoM length scale is not a unique function of the two twist angles. Figure 1c shows all possible twist-angle combinations for MoM-1 with color-scale representing the DOS maximum, with 1D cuts plotted in fig. 1d, corresponding to the scattered points in fig. 1c. The DOS exhibits a large value, comparable to tBLG, for a range of twist-angle configurations. In these regions the electronic band structure is characterized by flat bands at the charge-neutrality point (CNP) despite the lack of a band gap (figs. 1e-g show an example at $\theta_{12} = 2.9°$, $\theta_{23} = 1.4°$), signaling enhanced electronic correlation. The twist angles that exhibit flat bands (figs. 1c-d) agrees with the possible twist angles of the experimental devices considering layer migration and twist-angle inhomogeneity (see SI for more details).

In fig. 2a, we show the measured resistance $R$ (in $\Omega$) of the MoM-1 as a function of temperature spanning the entire range of band occupancy on the electron and hole sides. To describe the band occupancy, in the following we use the definition of "filling factor" $\nu = n/(n_s/4)$, where $n$ is the carrier density controlled by the silicon back gate and $n_s$ is the carrier density corresponding to four charge carriers per MoM unit cell (full-occupancy). Resistance peaks are observed at all even fillings, similar to twisted-bilayer systems but comparably much lower in amplitude. To demonstrate the temperature dependence of each filling, we plot the measured conductance as a function of carrier density at different temperatures (fig. 2c), and as a function of temperature at each signature filling factor (fig. 2e). We also present a control study of MoM-2 in figs. 2b, d, f. Here, $n_s = 6.22 \times 10^{10}$ cm$^{-2}$ ($8.37 \times 10^{10}$ cm$^{-2}$) is extracted from the resistance peak positions for the MoM-1 (MoM-2). For both studies, the $\nu$ = even integer peaks are clearly identifiable, with amplitude orders of magnitude larger than what can be expected from universal mesoscopic noise[33–35] and from the characterized noise level of the measurement system (see SI for details). At $v = -4$ and $v = 4$ for both MoM lattices, the measured resistance is similar to that of the charge neutrality point ($v = 0$) and displays a very weak metallic temperature dependence (more conducting at lower temperature), distinct from previously reported highly insulating gap states at $v = -4$ and 4 in the tBLG[2–4,36] and twisted double bilayer graphene systems[37,38]. This observation is consistent with our electronic structure result that tTLG is gapless and provides the first experimental evidence that correlated transport behavior can arise from non-isolated flat electronic bands.

In contrast to the semi-metallic behavior of the states at full-occupancy, the states at half-filling for MoM-1 exhibit an insulator-like temperature dependence (fig. 2c), with an extracted activation gap of 0.168 meV at $v = -2$ states (see SI for details), consistent with the observed metal-insulator transition near ~ 3 K. The definition of an insulator here is in the context of the temperature dependence only, as the absolute value of the measured resistance at half-filling is in fact lower than that of semi-metallic full-occupancy and zero-occupancy states. Around the $v = -2$ state, the measured resistance is observed to be zero below ~3K. The zero-resistance states are particle-hole asymmetric. Two zero-resistance domes are observed (corresponding to fig. 2) near $v = -2$, while two non-zero resistance dips are found near $v = 2$ filling (labeled as $v = 2^-$ and $v = 2^+$), as well as at both half-filling states of MoM-2 with strong metallic temperature dependence (fig. 3c). The widths of all resistance peaks including the CNP in each case are on the order of $10^{10}$ cm$^{-2}$, suggesting state-of-the-art device quality necessary for resolving the MoM correlated states.

The low carrier density of these resistance peaks corresponds to the filling of the tTLG MoM supercell (instead of the bilayer moiré supercells from either the 1-2 layer pair or the 2-3 layer pair in our tTLG device), which is two orders of magnitude larger in area (see SI for details) compared to the previously-reported tBLG moiré supercell. The corresponding possible angle configurations (fig. 1c and SI) are consistent with our experimentally-targeted value, considering unavoidable layer migration during the assembly process[5], reported to cause angle misalignment to up to ~ 1° in tBLG.

Due to the incommensurate nature of tTLG MoM superlattice, the small local twist-angle inhomogeneity in a completed tTLG stack (reported to be ~0.2° in tBLG[3,39]) can also result in a large spatial variation in the MoM unit cell area, making the observation of MoM correlated states extremely challenging. In MoM-2, which is found a few μm away on the same tTLG stack as MoM-1, even-filling states are reproduced and the metallic-like states are also observed near the half-filling. However, these states do not evolve into zero-resistance states at lower temperatures. Our observation is similar to the magic-angle continuum previously reported in twisted bilayer WSe$_2$ devices[28], where superconducting-like states are extremely sensitive to a slight change in twist angle while correlated insulator states are more robust.

In fig. 3a, we show the measured four-probe resistance near $\nu = -2$ for MoM-1 at zero magnetic field, as a function of carrier density and temperature. At a carrier density around $3.11 \times 10^{10}$ cm$^{-2}$, the resistance first decreases (corresponding to metallic behavior) from 20 K to 4 K, and then increases (insulating behavior) from 4 K down to 10 mK. We identify this change in behavior at 4 K as a correlated insulator state corresponding to half-filling, similar to that in tBLG[2,4]. Two zero-resistance domes are also found on either side of the half-filling insulating state below the same critical temperature, $T_c \sim 4$ K, which implies that correlated phases emerge below this critical temperature in our device. The span of the zero-resistance domes, as well as the carrier density at which each state was found, is about two orders of magnitude smaller than those in all the previously reported twisted bilayer systems. In fig. 3b, we show the resistance as a function of temperature, and in fig. 3c the 4-probe differential resistance as a function of carrier density and applied DC bias at $T = 3$ K. Zero differential resistance states exist in two domes symmetrically with respect to the zero bias, which is typically expected from a 2D superconductor. The critical current $J_c$ of the zero-resistance state at the higher doping is also visibly smaller, consistent with the observation in fig. 2a. The *I-V* characteristics (discussed also later in connection to fig. 4c) imply a critical current $J_c \sim 200$ nA even at 3 K, several times higher than the value previously reported for tBLG[2], $J_c \sim 50$ nA at 0.07 K. Moreover, as the temperature increases above $T_c$, the *I-V* characteristics (fig. 2d) evolve from cubic-like dependence (superconducting state) to linear dependence (normal state), consistent with the BKT model of 2D superconductivity[40]. The power-law of each *I-V* curve taken at different temperatures indicates a BKT transition temperature of $T_{BKT} \sim 3.4$ K (fig. 2d inset). The reported transport signatures (figs. 3 a-d) represent typical behavior expected from a 2D superconductor, similar to previously reported tBLG following the same well-developed experimental protocol. The observed width of the transitions (in temperature and *I-V*) are also consistent with a spatially varying superconducting order due to angle-inhomogeneity (see SI for detail).

However, additional experiments are needed in order to provide independent evidence for phase coherence (i.e., via gate-defined Josephson junctions) and Meissner-like effect, to confirm the superconducting nature of the observed zero-resistance states. Previous studies on tBLG show that critical current exhibits a weak oscillatory dependence[2] on the value of the in-plane magnetic field, as a result of accidental Josephson junction formation between co-existing superconducting ($v = 2$) and insulating ($v = 4$) domains due to twist-angle inhomogeneity. However, the $v = 4$ full-occupancy state in tTLG is semi-metallic and therefore a sufficiently-weak link for accidental

Josephson junction formation is fundamentally absent, as we have experimentally confirmed. Future demonstration of a similar "Fraunhofer pattern" in tTLG therefore may require a tTLG-hBN-tTLG vertical tunnel junction, which is beyond the scope of the present work.

Magneto transport studies are commonly used to characterize moiré length scales by tracing the origins of satellite Landau fan diagrams. Even in the absence of angle inhomogeneity, the MoM pattern is incommensurate by nature and a continuum of length scales coexists, resulting into a plethora of complicated magneto transport behavior that cannot be traced back to a definitive origin. To demonstrate these major differences from tBLG, we show magneto-transport data for the tTLG MoM-1 system (fig. 4). Due to the van der Pauw device geometry, the transport data is symmetrized to isolate the longitudinal resistances. The distances between each fan of $\nu =$ even integer fillings are as narrow as $\sim 10^{10}$ cm$^{-2}$, comparable to the narrowest features resolvable even in the highest quality graphene devices[41]. The amplitude of the observed transport features is also several orders of magnitude larger than the mesoscopic and system noise level. As expected from the extremely small density span between our $\nu = $ -4, 0, 4 insulator states, the magneto-transport in our tTLG device (figs. 4a, b) exhibits several sets of nearly overlapping Hofstadter butterfly patterns due to several sets of Landau fan diagrams in extreme proximity with each other[42]. In addition to the experimentally limited resolution of $10^{10}$ cm$^{-2}$, the indistinguishable Landau fans are also a signature of the tTLG systems due to their incommensurate nature even in the continuum limit. In tBLG systems, there always exists a dominant length scale: the bilayer moiré length. While the bilayer moiré pattern is not strictly periodic, it corresponds to the same repeating motif. In contrast, in tTLG, the interference between the two different misaligned bilayer moiré patterns can result in multiple MoM harmonics with comparable length scales at certain twist angles, leading to the lack of an apparent repeating pattern (see SI for details). Moreover, in tTLG, a small continuous change in the twist angle does not necessarily correspond to a smooth change in the length scale. With a slight change in the local twist angle, the MoM supercell area can be drastically different, as demonstrated in our control study comparing two MoM systems (see SI for details). Despite the complications, a zero-resistance dome can be identified (dashed line in fig. 4b) up to ~ 500 mT.

In conclusion, we have constructed an MoM superlattice in a novel twisted trilayer graphene architecture. The electronic band structure consists of a non-isolated zero-energy flat band, which we experimentally confirmed with observed semi-metallic resistance peaks at full-occupancy of the MoM supercell. Despite this major difference, correlated insulating states are found at half-filling, near which we also report zero-resistance states with transport signature typically expected from a 2D superconductor, at an unprecedently-low carrier density and a comparably high critical temperature of $T_{BKT}$ ~ 3.4 K. This demonstrates that an isolated flat band may not be required for the emergence of moiré correlated behaviors. While further experimental and theoretical studies on this new material platform are necessary to confirm the existence of superconductivity, our system exhibits both some similarities with the phase diagram in tBLG and transport properties distinct from tBLG, which shed new light on the microscopic origin of moiré correlated states.

**Competing interests**

The authors declare no competing interests.

**Additional information**

**Supplementary information** is available for this paper at [URL inserted by publisher]

**Correspondence and requests for materials** should be addressed to K.W. (kewang@umn.edu).

**Data availability**

All data needed to evaluate the conclusions in the paper are available from the corresponding author upon reasonable request.

**Code availability**

All relevant codes needed to evaluate the conclusions in the paper are available from the corresponding author upon reasonable request.

**Acknowledgements**

We thank Philip Kim, Allen Goldman, Paul Crowell and Hyobin Yoo for helpful experimental discussions, and Andrey Chubukov, Rafael Fernandes, Boris Shklovski, Dmitry Chichinadze, Laura Classen, and Daniel Massatt for helpful theoretical discussion. This work was supported in part by ARO MURI Award W911NF-14-0247and NSF DMREF Award 1922165. Z.Z and S.C. are supported by the STC Center for Integrated Quantum Materials, NSF Grant No. DMR-1231319. Nanofabrication was conducted in the Minnesota Nano Center, which is supported by the National Science Foundation through the National Nano Coordinated Infrastructure Network, Award Number NNCI -1542202.

**Contributions**

K-T. T., X. Z. and Y. L. performed the experiments, K-T. T. and X. Z. analyzed the data and fabricated the devices, K.W. conceived the device architecture and the experiment, Z. Z., S. C., E. K., and M. L. conceived, performed and analyzed the atomic simulation and band calculation.

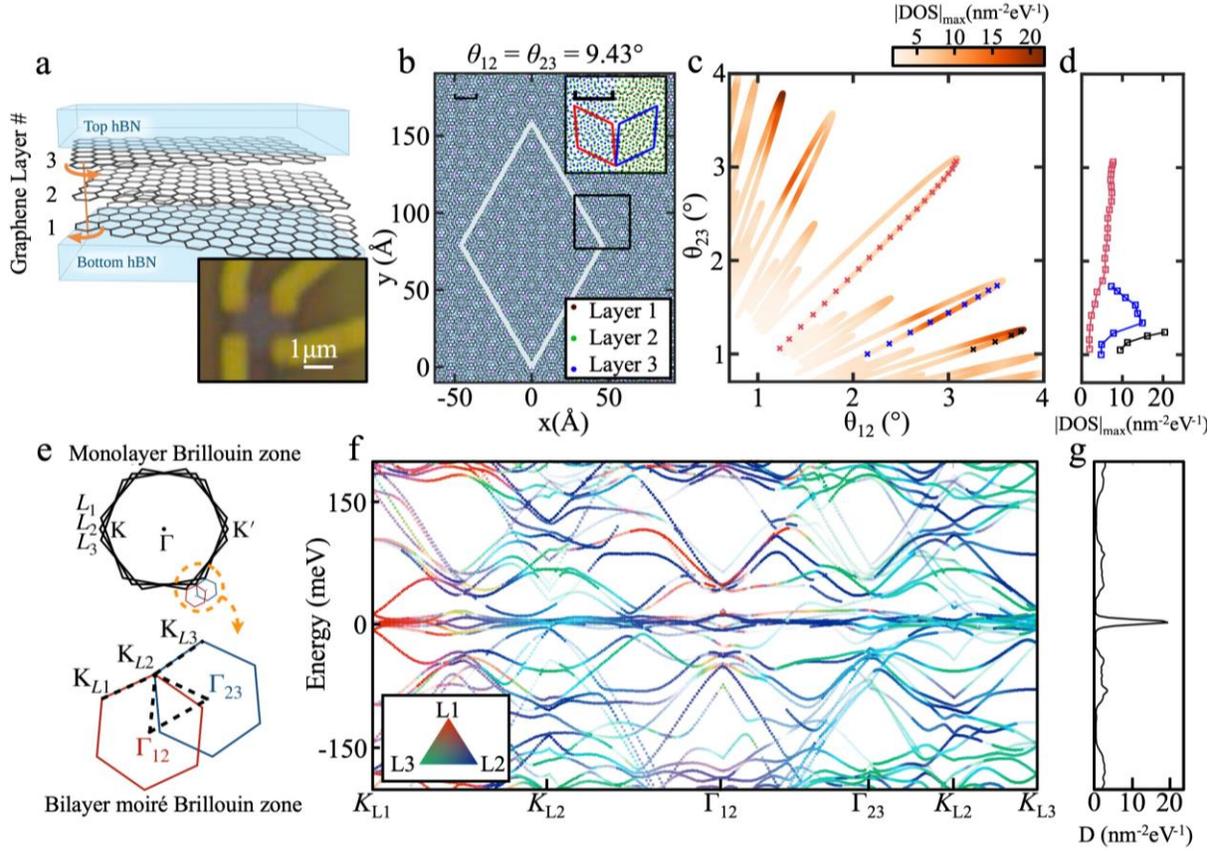

**Figure 1. Moiré of Moiré (MoM) Superlattices in Twisted Trilayer Graphene (tTLG).** (a) Schematic of the hBN-encapsulated tTLG device. Three individual pieces of monolayer graphene are transferred on top of each other, consecutively twisted by an angle with respect to the previous layer. (Inset) Optical microscope image of the van der Pauw device under study. (b) Atomic positions in tTLG at a set of commensurate angles $\theta_{12} = \theta_{23} = 9.43°$ demonstrating the MoM superlattice. Light green parallelogram: MoM supercell. (Inset) Zoom-in of the area inside the black box showing the bilayer moiré pattern. The left shows only layers 1 and 2; the right shows only layers 2 and 3; the red and blue parallelograms show the bilayer moiré supercells between layers 1 and 2 and layers 2 and 3 respectively. Both black scale bars indicate the length scale of 10 Å. (c) Possible twist angle combinations for MoM area $A_1$= 6438 nm$^2$ (MOM-1). Color scale represents the density of state maximum. Red, yellow and black dashed line denote(d) 1D cuts along twist angle configurations near $\theta_{12} = \theta_{23}$, $\theta_{12} = 2\theta_{23}$ and $\theta_{12} = 3\theta_{23}$, respectively. (e) Monolayer Brillouin zone of the three individual layers (top) and the bilayer moiré Brillouin zone formed between adjacent bilayer pairs (bottom). High symmetry points are labelled by letters. K and K′ and related by time reversal symmetry. (f) Band structure at $\theta_{12} = 2.9°$, $\theta_{23} = 1.4°$ along the high symmetry line indicated by the dashed line in (e). Colors indicate the projected weight onto the state at the center momentum in each layer. Red: layer 1; blue: layer 2; green: layer 3. The color scale is shown in the triangular colormap, with the corners of the triangles showing weight purely from a single layer. (g) Normalized density of states corresponding to band structure shown in (f), with zero-energy peak value comparable to that of tBLG.

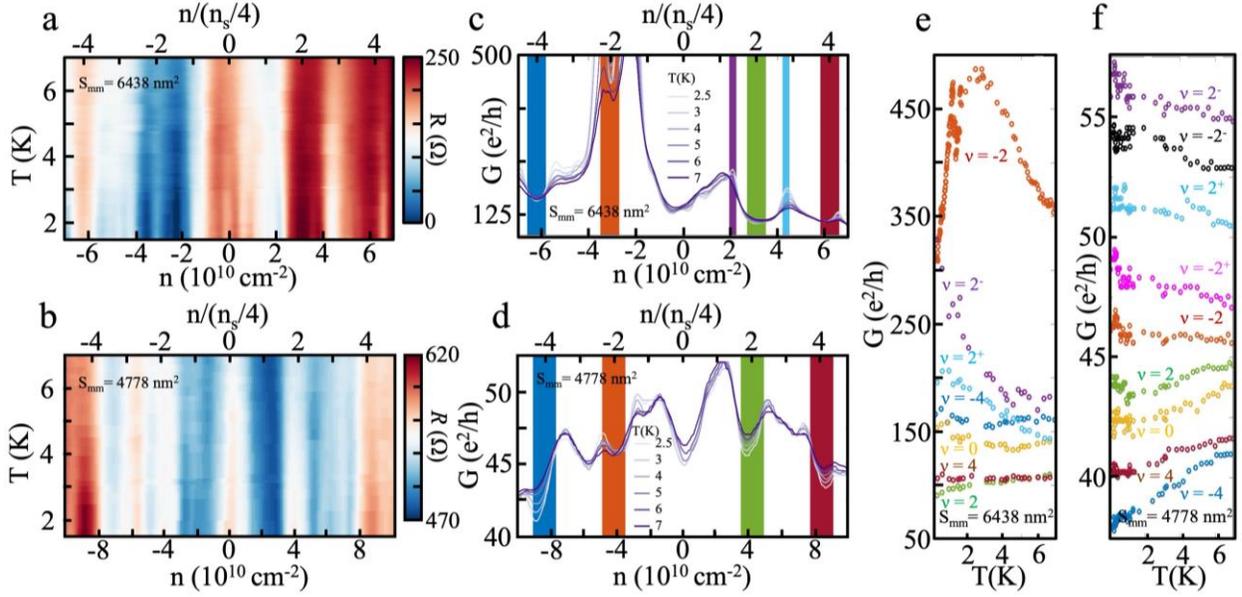

**Figure 2. Insulating and Semi-metallic Behavior at $\nu$ = even Integer Fillings of Two MoM Superlattices.** (a) Measured 4-probe resistance of MoM-1 (unit cell size = 6438 nm$^2$) as a function of temperature and carrier density, resistance peaks are observed at all $\nu = n/(n_s/4)$ = even integer fillings in which $n$ is the carrier density controlled by the silicon back gate, and $n_s = 6.22\times10^{10}$ cm$^{-2}$ is the carrier density corresponding to four charge carriers per MoM unit cell (full-occupancy). (b) Measured four-probe resistance as a function of temperature and carrier density of the controlled transport study of MoM-2 (unit cell size $A_2$= 4778 nm$^2$), similar to (a). (c) Line traces of four-probe conductance as a function of carrier density taken at different temperatures of MoM-1 with superconductivity signature. (d) Line traces of four-probe conductance as a function of carrier density taken at different temperatures of MoM-2. (e)-(f) Conductance of sample with superconductivity signature (MoM-1) and control transport study (MoM-2) at each even filling as a function of temperature. The curves in (f) are offset with respect to each other to clearly show their individual trends.

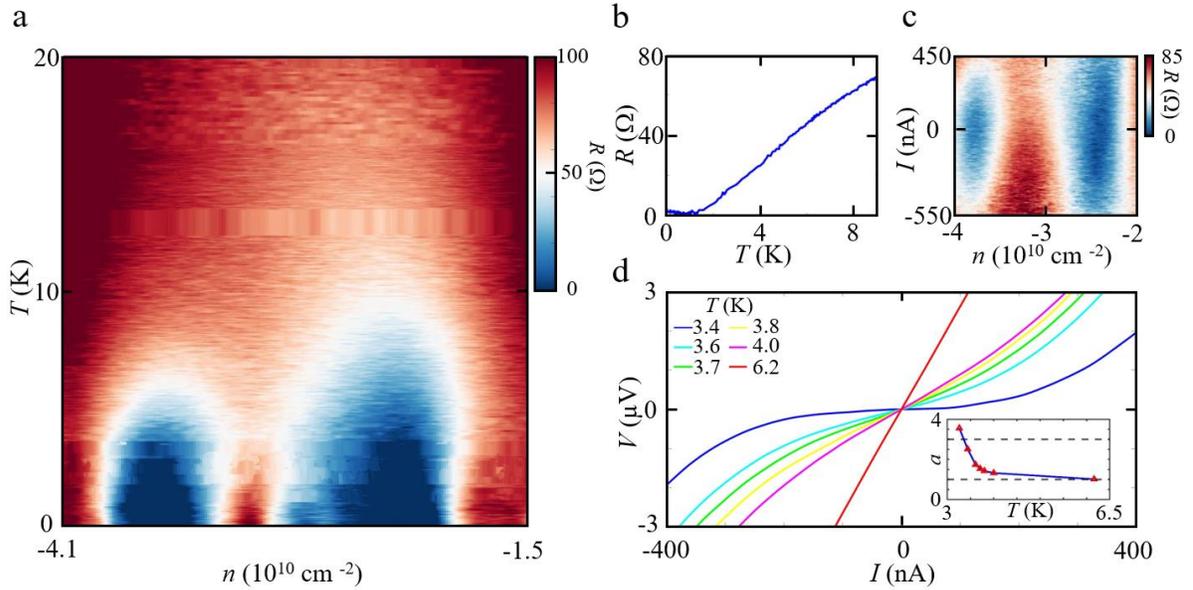

**Figure 3. Temperature and Bias Characteristics of Superconducting-like Zero-resistance States.** (a) Measured 4-probe resistance of MoM-1 as a function of carrier density and temperature. At half-filling of the MoM superlattice ($n \sim -3.22 \times 10^{10}$ cm$^{-2}$), correlated insulating behavior is observed with two adjacent zero-resistance domes. (b) Differential resistance as a function of temperature near the center of the right zero-resistance dome. (c) Differential resistance as a function of carrier density and DC bias offset at 3 K. Zero resistance states symmetric with respect to zero bias exist up to $J_c \sim 200$ nA. (d) I-V characteristics at different temperatures, near the center of the right zero-resistance dome. The I-V curve power law exponent goes from 1 to 3 as temperature decreases, consistent with the BKT model for a typical 2D superconductor. (Inset) The temperature dependence of the power law exponent $\alpha$, from which a critical temperature $T_{BKT} \sim 3.4$ K is extracted.

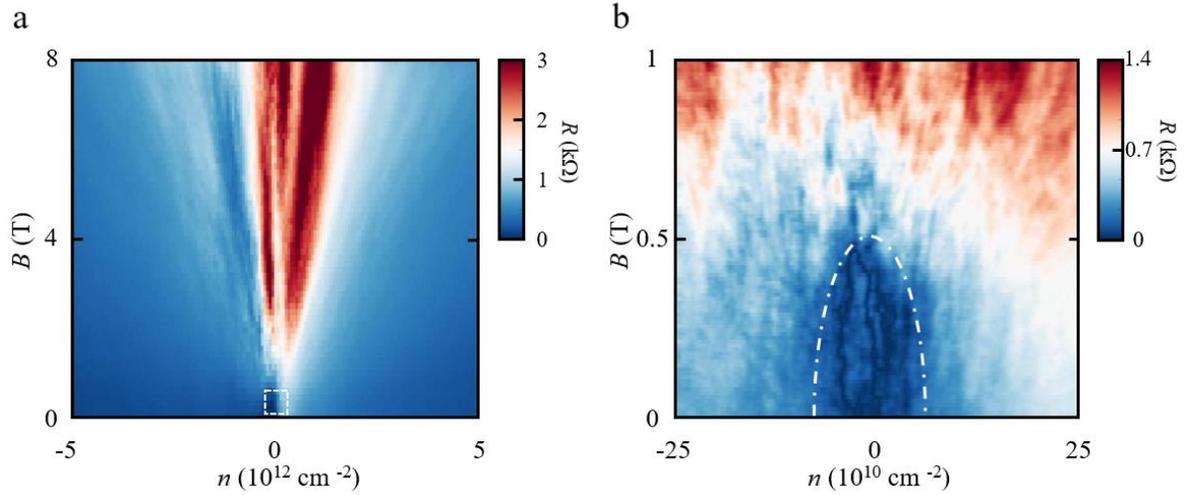

**Figure 4. Magneto-Transport at Larger Carrier Density Span.** (a) Resistance (*R*, in kΩ) of MoM-1 as a function of magnetic field (*B*, in *T*) and a large range of carrier density (*n*, in $10^{12}$ cm$^{-2}$). Zero resistance states are only observed in a very small density range near zero carrier density. (b) Magneto-transport data in a smaller range of carrier density, displaying a very complicated Hofstadter butterfly pattern, in which zero resistance states exist up to ~ 500 mT.